\documentstyle[12pt,bezier]{article}
\begin{document}
\baselineskip 18pt
\def\today{\ifcase\month\or
 January\or February\or March\or April\or May\or June\or
 July\or August\or September\or October\or November\or December\fi
 \space\number\day, \number\year}
\def\thebibliography#1{\section*{References\markboth
 {References}{References}}\list
 {[\arabic{enumi}]}{\settowidth\labelwidth{[#1]}
 \leftmargin\labelwidth
 \advance\leftmargin\labelsep
 \usecounter{enumi}}
 \def\newblock{\hskip .11em plus .33em minus .07em}
 \sloppy
 \sfcode`\.=1000\relax}
\let\endthebibliography=\endlist
\def\gsim{~{\rlap{\lower 3.5pt\hbox{$\mathchar\sim$}}\raise 1pt\hbox{$>$}}\,}
\def\lsim{~{\rlap{\lower 3.5pt\hbox{$\mathchar\sim$}}\raise 1pt\hbox{$<$}}\,}
\def\epem{e^+e^-}
\def\r2{\sqrt 2}
\def\mqu#1{m_{u#1}}
\def\mqd#1{m_{d#1}}
\def\m#1{{\tilde m}_#1}
\def\mH{m_H}
\def\mw#1{m_{\omega #1}}
\def\mx#1{m_{\chi #1}}
\def\mwi{m_{\omega i}}
\def\mxj{m_{\chi j}}
\def\ri{\rm i}
\def\rj{\rm j}
\def\rk{\rm k}
\begin{titlepage}
\hspace*{10.0cm}ICRR-Report-448-99-6 
   
\hspace*{10.0cm}OCHA-PP-132

\hspace*{10.0cm}KURE-PP/99-01
\  \
\vskip 0.5 true cm 
\begin{center}
{\large {\bf CP-odd $WWZ$ couplings induced by vector-like quarks}}  
\vskip 2.0 true cm
\renewcommand{\thefootnote}
{\fnsymbol{footnote}}
Eri Asakawa$^1$, Miho Marui$^2$, Noriyuki Oshimo$^3$, 
Tomomi Saito$^1$\footnote{
Present address:  Konami Co. LTD., 
Tokyo 101-0051, Japan.}, and \\
Akio Sugamoto$^1$ 
\\
\vskip 0.5 true cm 
{\it $^1$Department of Physics {\rm and} 
Graduate School of Humanities and Sciences }  \\
{\it Ochanomizu University, Otsuka 2-1-1, 
Bunkyo-ku, Tokyo 112-8610, Japan}  \\
{\it $^2$Faculty of Social Information Science}  \\
{\it Kure University, Gouhara 2411-26, Kure, Hiroshima 724-0792, Japan}  \\
{\it $^3$Institute for Cosmic Ray Research}  \\
{\it University of Tokyo, Midori-cho 3-2-1, 
Tanashi, Tokyo 188-8502, Japan}  \\
\end{center}

\vskip 3.0 true cm

\centerline{\bf Abstract}
\medskip
     A minimal extension of the standard model includes extra   
quarks with charges 2/3 and/or $-1/3$, whose left-handed and right-handed 
components are both SU(2) singlets.   
This model predicts new interactions of   
flavor-changing neutral current at the tree level, which also 
violate CP invariance.  
We study CP-odd anomalous couplings for the gauge bosons $W$, $W$, and $Z$ 
induced by the new interactions at the one-loop level.  
These couplings become nonnegligible only if 
both an up-type and a down-type extra quarks are incorporated.   
Their form factors are estimated to be maximally of order $10^{-5}$.     
Such magnitudes are larger than those predicted in the standard model,  
though smaller than those in certain other models.  
\medskip

\end{titlepage}

\newpage 
\section{Introduction} 

    Experiments at LEP2 and planned $\epem$ linear colliders 
can directly probe gauge-boson self-interactions, an aspect 
of the standard model (SM) characteristic of its non-Abelian nature.  
Their precise measurements serve a detailed examination of the SM, 
whose predictions have been well studied including quantum corrections.   
If some deviations from the SM predictions are found, 
the SM will have to be extended.  
For this task, studying peculiar features of various models is indispensable.  
Theoretical analyses therefore have been made 
on the gauge-boson self-interactions in the extensions of 
the SM \cite{hagiwara,schild}.   
In particular, CP-even couplings for the gauge bosons $W$, $W$, and $Z$  
have been studied in the two-Higgs-doublet model \cite{rizzo}, the model 
with Majorana neutrinos \cite{marui}, the supersymmetric 
model \cite{lahanas}, and so on.   
The CP-odd $WWZ$ couplings have also recently been discussed in the 
supersymmetric model \cite{kitahara}.   

     One of minimal extensions of the SM is the vector-like quark model (VQM).  
This model includes extra quarks with charges 2/3 and/or $-1/3$, 
whose left-handed components, as well as right-handed ones, 
are singlets under SU(2).  
Although many features of the SM are not significantly modified, 
new sources of CP violation and flavor-changing neutral current (FCNC) 
are incorporated.  
Therefore, their effects on the $K$-meson and $B$-meson systems 
have been studied extensively \cite{branco}.  
It has been also argued \cite{uesugi} that baryon asymmetry of the universe 
could be attributed to these new sources of CP violation.  

     In this paper, we study the effects of the VQM on CP-odd couplings  
for the $WWZ$ vertex.   
This model predicts that the $Z$ boson couples to quarks of 
different generations, causing interactions of FCNC at the tree level.  
For a pair of light ordinary quarks, these interactions should be suppressed 
from experimental results.    
However, the $t$ quark and the extra up-type quark $U$ could have a sizable 
coupling with the $Z$ boson.  
Since these interactions of FCNC also violate CP invariance, nonnegligible 
CP-odd couplings for the $WWZ$ vertex may be induced at the one-loop level.    
On the other hand, the SM does not contain the CP-odd $WWZ$ couplings  
at the tree level nor the one-loop level.    
The supersymmetric model predicts them at the one-loop level \cite{kitahara}.  
The CP-odd $WWZ$ couplings at the one-loop level could become a window  
for physics beyond the SM.  
It will be shown that form factors for the couplings in the VQM  
can be evaluated without making many assumptions on mixings among quarks.  
The form factors are nonnegligible at the one-loop level, 
though the possible maximal magnitudes are smaller than those in the  
supersymmetric model.  

     This paper is organized as follows.  In sect. 2 we briefly summarize 
the model.  In sect. 3 the CP-odd form factors are obtained and   
numerical analyses are performed.  Summary is contained in sect. 4.  

\section{Model}

     The quark sector is enlarged to have extra quarks 
whose transformation properties are given by $(3, 1, 2/3)$ or $(3,1,-1/3)$ 
for the SU(3)$\times$SU(2)$\times$U(1) gauge symmetry.  
Both the left-handed and right-handed components have the same properties.  
For definiteness, we assume the particle contents with one up-type 
and one down-type extra quarks.   
The quark masses are generated by Yukawa couplings and bare mass terms.   
The mass matrices are given by 4$\times$4 matrices, which are denoted 
by $M^u$ and $M^d$ respectively for up-type and down-type quarks.  
The mass eigenstates are obtained by diagonalizing the mass matrices as 
\begin{eqnarray}
      A_L^{u\dagger} M^uA^u_R &=& {\rm diag}(\mqu1,\mqu2,\mqu3,\mqu4),   \\
      A_L^{d\dagger} M^dA^d_R &=& {\rm diag}(\mqd1,\mqd2,\mqd3,\mqd4),   
\end{eqnarray}
where $A^u_L$, $A^u_R$, $A^d_L$, and $A^d_R$  are unitary matrices.  
We express the mass eigenstates by $u^a$ and $d^a$ ($a=1-4$),   
$a$ being the generation index, which may be also called as 
$(u,c,t,U)$ and $(d,s,b,D)$.    

     The interaction Lagrangian for the quarks with the $W$ boson is  
given by 
\begin{equation}
 {\cal L} = \frac{g}{\r2}\overline{u^a}V_{ab}\gamma^\mu               
                   \frac{1-\gamma_5}{2}d^bW_\mu^\dagger + {\rm h.c.}.    
\label{Wlagrangian}
\end{equation}
Here the $4\times 4$ matrix $V$ stands for an extended 
Cabibbo-Kobayashi-Maskawa matrix, which is defined by 
\begin{equation}
      V_{ab} = \sum_{i=1}^3(A_L^{u\dagger})_{ai}(A_L^d)_{ib}.  
\end{equation}
Note that $V$ is not unitary.  
The interaction Lagrangian for the quarks with the $Z$ boson is given by  
\begin{eqnarray}
 {\cal L}&=&-\frac{g}{\cos\theta_{\rm W}}
          \overline{u^a}\gamma^\mu
            \left( F_{Lab}^u\frac{1-\gamma_5}{2}
               +F_{Rab}^u\frac{1+\gamma_5}{2} \right) u^bZ_\mu   \nonumber \\
          & & -\frac{g}{\cos\theta_{\rm W}}
          \overline{d^a}\gamma^\mu
            \left( F_{Lab}^d\frac{1-\gamma_5}{2}
               +F_{Rab}^d\frac{1+\gamma_5}{2} \right) d^bZ_\mu,  
\label{Zlagrangian}  \\ 
  F_L^u &=& \frac{1}{2}VV^\dagger-\frac{2}{3}\sin^2\theta_W,  
 \quad F_R^u = -\frac{2}{3}\sin^2\theta_W,              \nonumber \\
     F_L^d &=& -\frac{1}{2}V^\dagger V
                      +\frac{1}{3}\sin^2\theta_W,  
 \quad F_R^d = \frac{1}{3}\sin^2\theta_W.    \nonumber  
\end{eqnarray}
Since $V$ is not a unitary matrix, off-diagonal elements of  
$F_L^u$ and $F_L^d$ become nonvanishing, leading to FCNC at the tree level.     
The Lagrangian in Eq. (\ref{Zlagrangian}), as well as that in 
Eq. (\ref{Wlagrangian}), can induce CP violation.  

\section{Form factors}

     For the pair production of $W$ bosons in $\epem$ annihilation,  
the trilinear gauge-boson interaction for $W$, $W$, and $Z$ 
is generally expressed as \cite{hagiwara}  
\begin{eqnarray}
{\cal L}_{eff} &=& g\cos\theta_{\rm W}\Gamma^{\mu\nu\lambda}
                   W_\mu^\dagger W_\nu Z_\lambda,  \\
\Gamma^{\mu\nu\lambda} &=& 
       f_{1} (p-\bar{p})^{\lambda} g^{\mu\nu} 
     + f_{2} \frac{1}{M_W^2}(p-\bar{p})^{\lambda} q^{\mu} q^{\nu} 
     + f_{3} ( q^{\mu} g^{\lambda\nu} - q^{\nu} g^{\lambda\mu} ) 
                      \nonumber \\ 
 & & + i f_{4} (q^{\mu} g^{\lambda\nu} + q^{\nu} g^{\lambda\mu} ) 
     + i f_{5} \varepsilon^{\mu\nu\lambda\rho} (p - \bar{p} )_{\rho} 
     + f_{6} \varepsilon^{\mu\nu\lambda\rho} q_{\rho}  
                      \nonumber \\
 & & + f_{7} \frac{1}{M_W^2}(p-\bar{p})^{\lambda}\varepsilon^{\mu\nu\rho\sigma}
                 q_{\rho} (p-\bar{p} )_{\sigma},   \nonumber 
\end{eqnarray}
where $p$ and $\bar p$ denote the outward momenta of the gauge bosons
$W^-$ and $W^+$, respectively, and $q$ the inward momentum of $Z$.   
The couplings with the form factors $f_1$, $f_2$, $f_3$, and $f_5$ 
are $CP$-even, while those with $f_4$, $f_6$, and $f_7$ are $CP$-odd.   
Both in the SM and in the VQM, only the form factors $f_1$ and $f_3$  
have non-vanishing values at the tree level.  

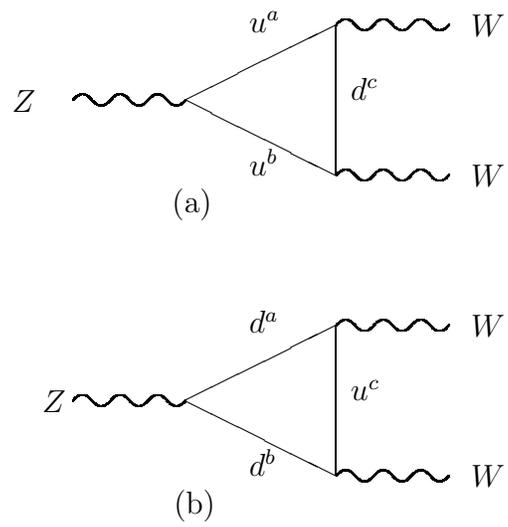
\begin{figure}
\begin{center}
   \setlength{\unitlength}{1mm}
   \begin{picture}(70,90)

\multiput(3,5)(0,40){2}{
  \multiput(5,15)(5,0){3}{\bezier{200}(0,0)(1.25,1.5)(2.5,0)}
  \multiput(7.5,15)(5,0){3}{\bezier{200}(0,0)(1.25,-1.5)(2.5,0)}
  \multiput(40,25)(5,0){3}{\bezier{200}(0,0)(1.25,1.5)(2.5,0)}
  \multiput(42.5,25)(5,0){3}{\bezier{200}(0,0)(1.25,-1.5)(2.5,0)}
  \multiput(40,5)(5,0){3}{\bezier{200}(0,0)(1.25,1.5)(2.5,0)}
  \multiput(42.5,5)(5,0){3}{\bezier{200}(0,0)(1.25,-1.5)(2.5,0)}
  \put(20,15){\line(2,1){20}}
  \put(20,15){\line(2,-1){20}}
  \put(40,25){\line(0,-1){20}}
  \put(58,23.5){$W$}
  \put(58,3.5){$W$}
}
  \put(0,58.5){$Z$}
  \put(4,18.5){$Z$}

\multiput(3,5)(0,40){1}{
     \put(28.5,64){$u^a$}
     \put(28.5,45){$u^b$}
     \put(42,55){$d^c$}
     \put(28.5,24){$d^a$}
     \put(28.5,5){$d^b$}
     \put(42,15){$u^c$}
}
     \put(21.25,45){(a)}
     \put(21.5,5){(b)}

    \end{picture}
\end{center}
\caption{Possible one-loop diagrams which induce $CP$-odd couplings 
for the $WWZ$ vertex.}
\label{fig:oneloop}
\end{figure}

     The $CP$-odd form factors receive contributions from 
the one-loop diagrams in which up-type quarks or down-type quarks 
couple to the $Z$ boson as shown in Fig.~\ref{fig:oneloop}.  
We obtain the form factors arising from the diagram 
in Fig. \ref{fig:oneloop}(a) as  
\begin{eqnarray} 
f_4 &=&\frac{-g^2}{64\pi^2\cos^2\theta_{\rm W}}
         \sum_{a=1}^4 \sum_{b=1}^4\sum_{c=1}^4     
              {\rm Im} \left[V_{ac}V^\ast_{bc}(VV^\dagger)_{ba}\right]  
               I_4(m_{ua},m_{ub},m_{dc}),  
\label{form4} \\  
f_6 &=&\frac{-g^2}{64\pi^2\cos^2\theta_{\rm W}}
          \sum_{a=1}^4 \sum_{b=1}^4\sum_{c=1}^4     
              {\rm Im} \left[V_{ac}V^\ast_{bc}(VV^\dagger)_{ba}\right]  
             I_6(m_{ua},m_{ub},m_{dc}),  
\label{form6} \\
f_7&=&0,    
\end{eqnarray}
where the functions $I_4$ and $I_6$ are defined by 
\begin{eqnarray} 
   I_4(m_{ua},m_{ub},m_{dc}) &=&  {} 
\label{int4}
\end{eqnarray}
\[  
    \int\int_D dxdy 
      \frac{M^2_W (1-x-y)^2(x-y) + (m_{ua}^2-m_{ub}^2)xy} 
 {-M^2_W (1-x-y)(x+y)-q^2xy+m^2_{ua}x+m^2_{ub}y+m^2_{dc}(1-x-y)-i\varepsilon}, 
\]           
\begin{eqnarray} 
    I_6(m_{ua},m_{ub},m_{dc}) &=& {} 
\label{int6}
\end{eqnarray}
\[  
      \int\int_D dxdy   
       \frac{M^2_W (1-x-y)(1+x+y)(x-y)-2q^2 xy(x-y)
                  -3(m_{ua}^2-m_{ub}^2)xy}
 {-M^2_W (1-x-y)(x+y)-q^2xy+m^2_{ua}x+m^2_{ub}y+m^2_{dc}(1-x-y)-i\varepsilon}.  
\]
The domain $D$ for integration is given by 
\begin{equation}
   x \geq 0, \quad y \geq 0, \quad x+y \leq 1.  \nonumber 
\end{equation}
The contributions of the diagram in Fig. \ref{fig:oneloop}(b) 
can be obtained similarly.  However, these contributions 
are negligible compared to those from Fig. \ref{fig:oneloop}(a) 
as shown later.  

     The form factors $f_4$ and $f_6$ in Eqs. (\ref{form4}) and 
(\ref{form6}) are expressed more simply by 
taking approximations for the quark masses.  
Since the quarks of the first two generations and the $b$ quark 
are much lighter than the $W$ boson, the integrands of  
$I_4$ and $I_6$ in Eqs. (\ref{int4}) and (\ref{int6}) 
are determined almost only by the $W$-boson mass $M_W$,  
the heavy quark masses $m_t$, $m_U$, $m_D$, and the momentum-squared 
$q^2$ of the $Z$ boson.  
We can safely neglect mass differences among the light quarks.  
Then, taking $m_u=m_c$ and $m_d=m_s=m_b$, 
the form factors are written by  
\begin{eqnarray} 
f_i &=&\frac{g^2}{32\pi^2\cos^2\theta_{\rm W}}CS_i \quad (i=4,6),  
 \label{nuformfactor} \\  
    C &=&  {\rm Im} \left[V_{34}V^\ast_{44}(VV^\dagger)_{43}\right], \nonumber \\ 
S_i &=&  I_i(m_u,m_t,m_d) - I_i(m_u,m_U,m_d) + I_i(m_t,m_U,m_d) \nonumber \\
    & & - I_i(m_u,m_t,m_D) + I_i(m_u,m_U,m_D) - I_i(m_t,m_U,m_D). \nonumber 
\label{eq:S}
\end{eqnarray}
Here, $C$ depends on mixing parameters for quarks, while  
$S_4$ and $S_6$ depend on mass parameters $m_U$, $m_D$ 
and an experimental parameter $q^2$.  
It is seen that $S_4$ and $S_6$ vanish if an equality 
$m_u=m_t$ is assumed.  
Correspondingly, the contributions to $f_4$ and $f_6$ from the diagram in 
Fig. \ref{fig:oneloop}(b) become negligible, since an equality $m_d=m_b$ 
holds to a good approximation.      

     Numerical analyses for the form factors are now in order.  
We first consider $S_4$ and $S_6$.  The integrals in Eqs. (\ref{int4}) 
and (\ref{int6}) are precisely evaluated by a numerical 
method \cite{kato}.  
In Figs. 2(a) and 2(b) the absolute values for the real and imaginary 
parts of $S_4$ and $S_6$ are shown 
as functions of $m_U$ for $m_U\geq 200$ GeV, taking $\sqrt{q^2}=200$ GeV 
and $m_D=$ 200 GeV (Fig. 2(a)), 500 GeV (Fig. 2(b)).    
Curves (i), (ii), (iii), and (iv) represent ${\rm Re}(S_4)$, 
${\rm Im}(S_4)$, ${\rm Re}(S_6)$, and ${\rm Im}(S_6)$, respectively.    
In wide ranges the imaginary parts do not vary much with $m_U$ and $m_D$, 
being Im$(S_4)\approx -0.3$ and Im$(S_6)\approx 0.9$.     
The real parts are in the ranges $|{\rm Re}(S_4)|\lsim 0.2$ 
and $|{\rm Re}(S_6)|\lsim 0.4$.   
The values of $S_4$ and $S_6$ are not much dependent on $m_D$ for 
$m_D\gsim 400$ GeV.    
In Fig. 3 the $\sqrt{q^2}$-dependencies of $S_4$ and $S_6$ 
are shown for $\sqrt{q^2}\geq 180$ GeV, 
taking $m_U=300$ GeV and $m_D=300$ GeV.  
Curves (i), (ii), (iii), and (iv) stand for the same objects as in Fig. 2.  
The absolute values decrease, as $\sqrt{q^2}$ increases.  

     Next we consider the magnitude of $C$.  It is seen from 
Eqs. (\ref{Wlagrangian}) and (\ref{Zlagrangian}) that 
$V_{34}$, $V^\ast_{44}$, and $(VV^\dagger)_{43}$ are related to  
the couplings of $tDW$, $UDW$, and $UtZ$, respectively.  
Experimental results available at present have not yet given much information 
on their values.     
However, $C$ is expressed in terms of the unitary matrices 
$A_L^u$ and $A_L^d$ as 
\begin{eqnarray}
 C &=& -{\rm Im}\biggl[(A_L^u)_{43}(A_L^{u\dagger}A_L^d)_{34}
             (A_L^{d\dagger}A_L^u)_{44}(A_L^{u\dagger})_{44}     
  -(A_L^u)_{43}(A_L^{u\dagger}A_L^d)_{34}
             (A_L^{d\dagger})_{44}|(A_L^u)_{44}|^2     \nonumber \\
  & & -|(A_L^u)_{43}|^2(A_L^d)_{44}
             (A_L^{d\dagger}A_L^u)_{44}(A_L^{u\dagger})_{44}\biggr].   
\end{eqnarray}
Since the mass of the $D$ quark is considered to be significantly 
larger than those of the $d$ and $s$ quarks, 
we may neglect mixings between the $D$ quark and these light quarks, 
taking $(A_L^d)_{i4}=(A_L^d)_{4i}=0$ ($i=1,2$). 
Then, $C$ is given by 
\begin{eqnarray}
C &=& -|(A_L^d)_{34}|^2{\rm Im}\left[(A_L^u)_{43}(A_L^{u\dagger})_{33}
             (A_L^u)_{34}(A_L^{u\dagger})_{44}\right].       
\end{eqnarray}
We can see that $C$ is proportional to $(A_L^d)_{34}$.   
If there is no mixing between the $D$ quark and the ordinary down-type 
quarks, the form factors $f_4$ and $f_6$ vanish.  
In order to have nonnegligible CP-odd couplings, the extra down-type quark 
$D$, as well as $U$, should exist and be mixed with other down-type quarks.   
The magnitude of $C$ is estimated to be at most of order 0.1, 
since $C$ contains the product of four different elements of a unitary matrix   
in addition to $|(A_L^d)_{34}|^2$.   
If the mixings of the quarks for the third and fourth generations 
are not suppressed, $|C|$ becomes maximal.   
However, the mixing for the down-type quarks would be suppressed,  
since the mass difference between the $b$ quark and 
the $D$ quark is still large.   
Therefore, we take the allowed range of $C$ for $|C|<0.01$ 
as a conservative constraint.  

     The form factors are written from Eq. (\ref{nuformfactor}) as 
\begin{equation} 
f_i = 1.7\times 10^{-3}CS_i \quad (i=4,6).  
\end{equation}
Taking into account the constraints on $C$ and $S_i$, we make an estimate  
\begin{equation}
   |f_4|, |f_6| < 1\times 10^{-5}.  
\end{equation}
Possible maximal values of this result are larger than the predictions in the SM, 
though smaller by two order of magnitude than those 
in the supersymmetric model \cite{kitahara}, 
where the CP-odd form factors are also induced at the one-loop level.  
Assuming a maximal value, a total of more than $10^{10}$ pairs of 
$W$ bosons would be necessary to detect the form factors.   
It seems to be difficult to achieve such a number of events 
in near-future experiments.   

\section{Summary}

     We have discussed CP-odd couplings for the $WWZ$ vertex 
within the framework of the VQM.  These couplings could be  
sizably induced through the one-loop diagram in which 
the $Z$ boson couples to the up-type quarks.  
Both up-type and down-type extra quarks are necessary to have 
nonnegligible form factors.    
Their possible maximal magnitudes have been estimated without assuming 
a detailed structure for the quark mixings, giving at most     
of order $10^{-5}$.  
These magnitudes are larger than the predictions by the SM but smaller than 
those by the supersymmetric model.    
The VQM does not yield CP-odd $WWZ$ coupligs which can be detected 
experimentally in the near future.

\section*{Acknowledgments}
 
This  work is supported in part by the Grant-in-Aid for Scientific
Research (No. 08640357, No. 08640400) and the Grant-in-Aid for 
Scientific Research on Priority Areas (Physics of $CP$ Violation, 
No. 10140208) from the Ministry of Education, Science and
Culture, Japan.

\newpage 
 
%


\pagebreak

\begin{figure}
\vspace{12cm}
\includegraphics{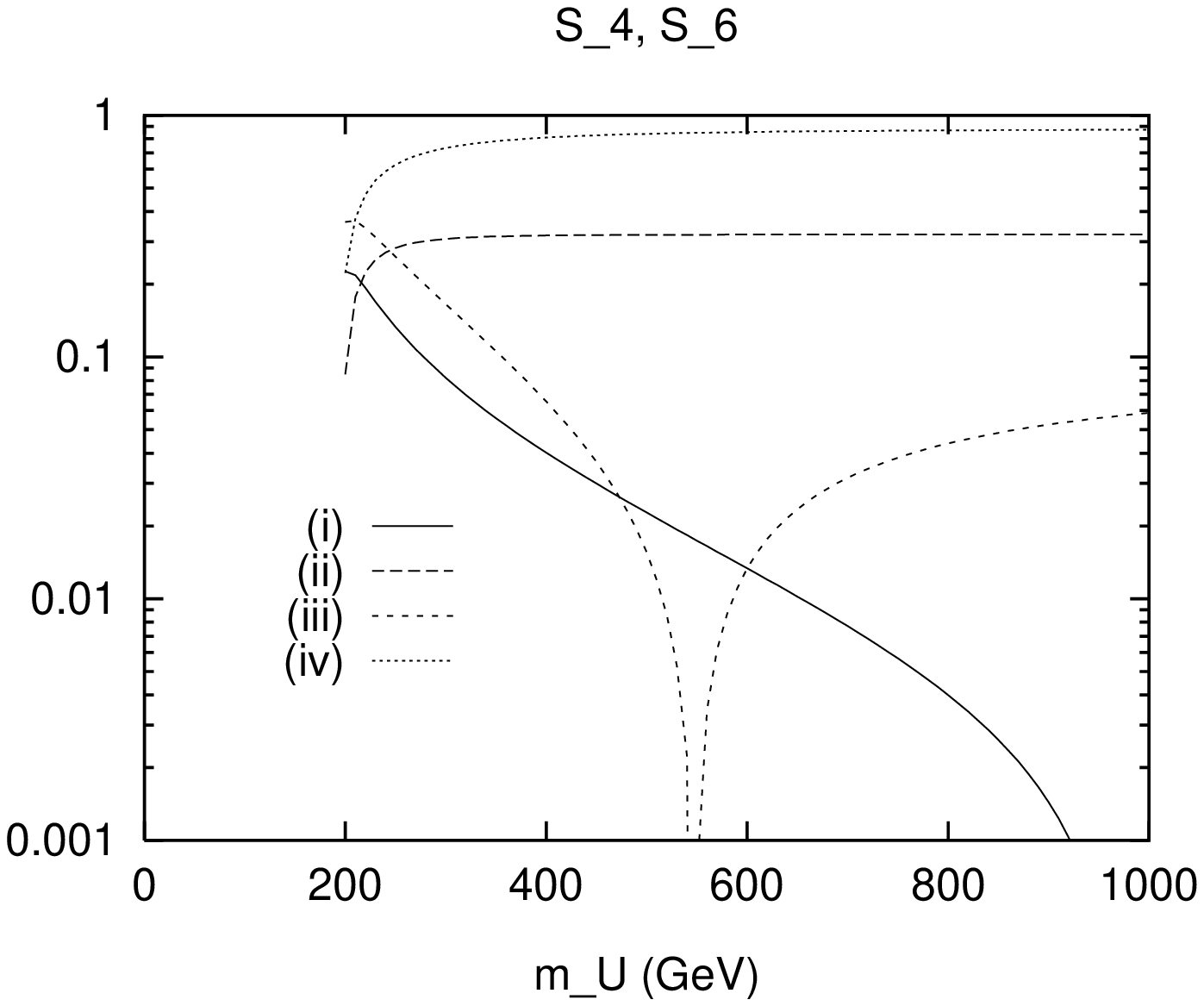}
\begin{center}
Fig. 2(a)
\end{center}
\end{figure}

\pagebreak

\begin{figure}
\vspace{12cm}
\includegraphics{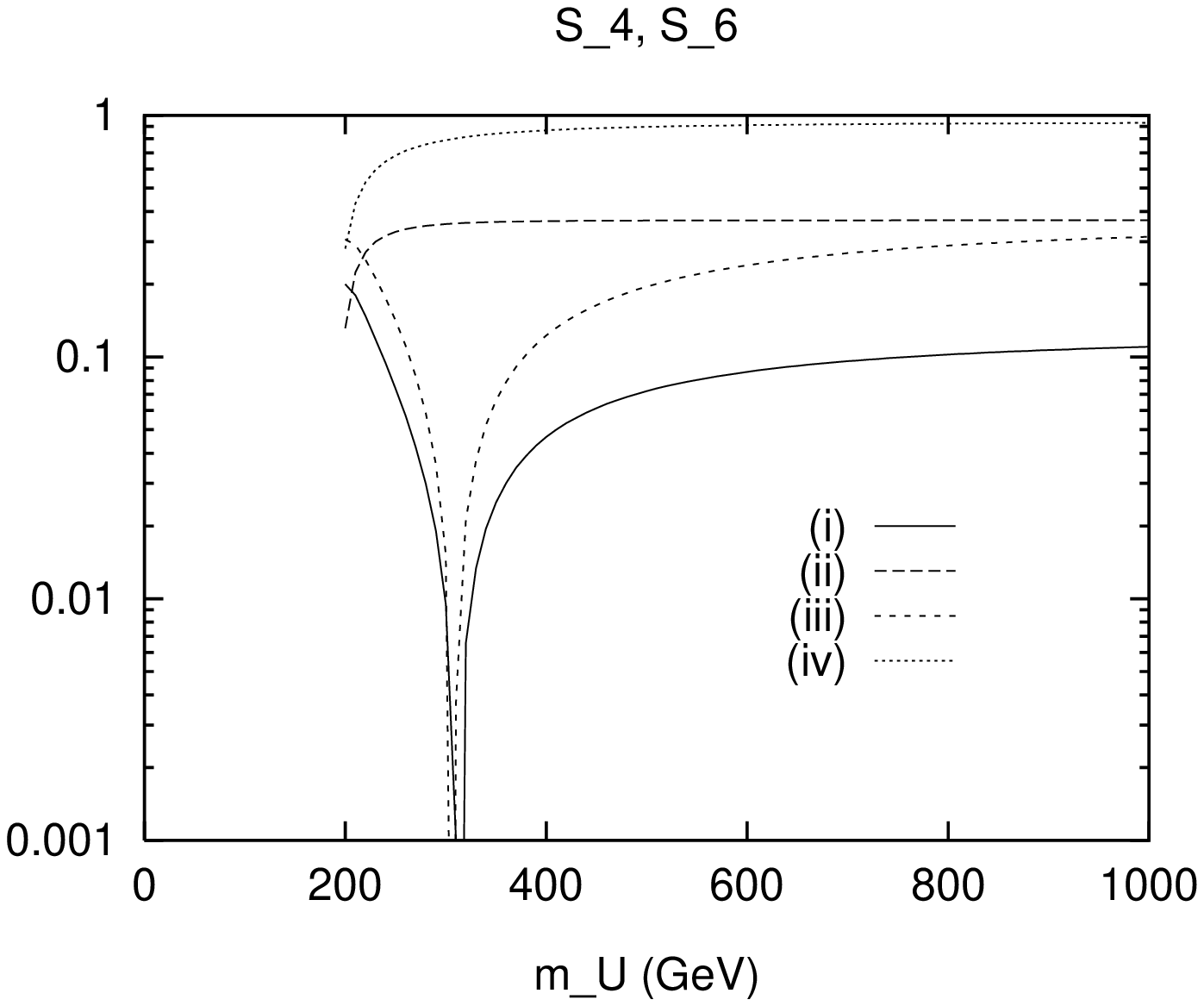}
\begin{center}
Fig. 2(b)
\end{center}
\caption{The absolute values of the real and imaginary 
parts of $S_4$ and $S_6$ as functions of $m_U$ 
at $\protect \sqrt{q^2}=200$ GeV.   
Four curves (i)--(iv) correspond to  
${\rm Re}(S_4)$, ${\rm Im}(S_4)$, ${\rm Re}(S_6)$, ${\rm Im}(S_6)$.    
(a) $m_D=200$ GeV,  (b) $m_D=500$ GeV.} 
\end{figure}

\pagebreak

\begin{figure}
\vspace{12cm}
\includegraphics{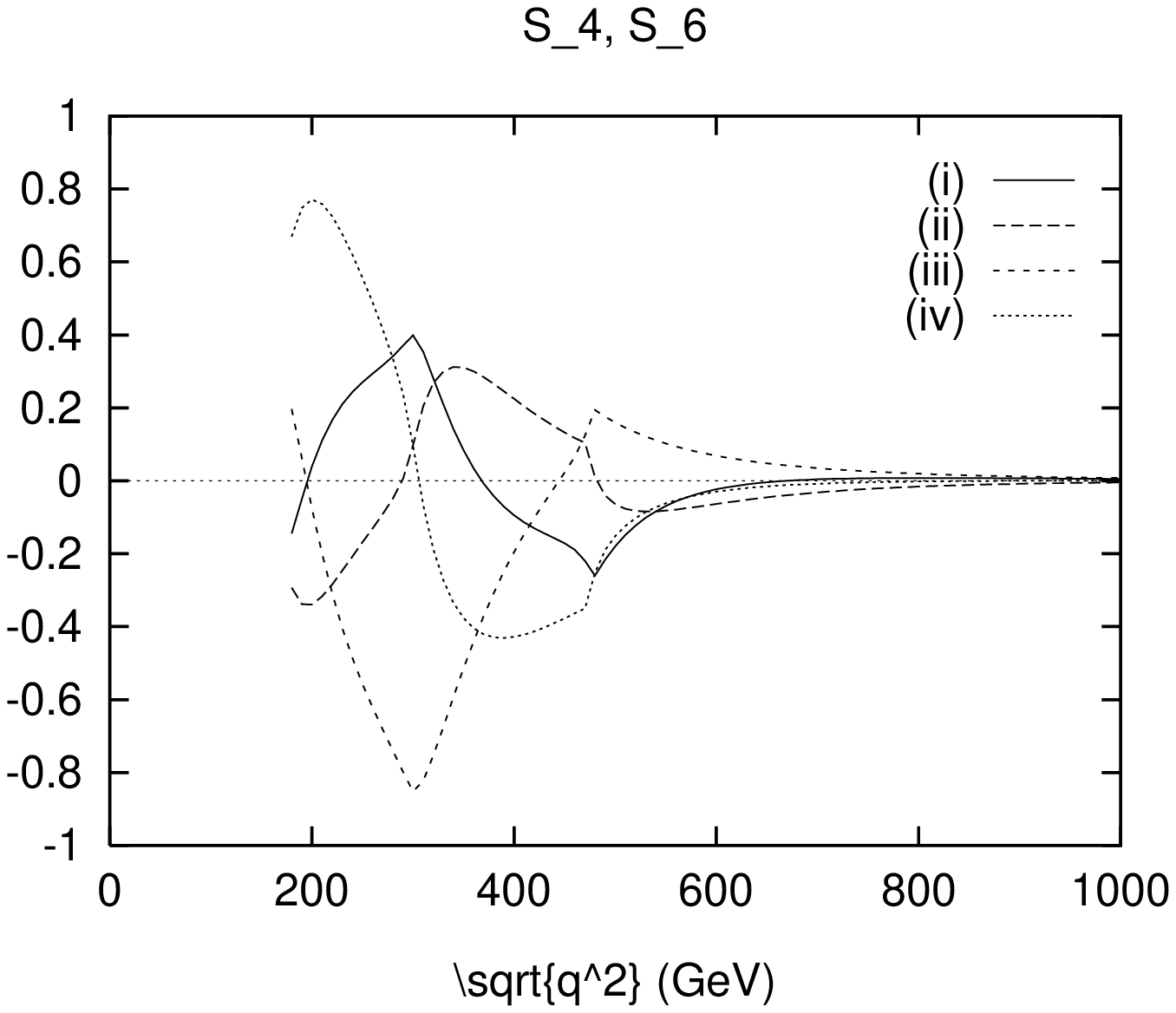}
\begin{center}
Fig. 3
\end{center}
\caption{The values of the real and imaginary parts of 
$S_4$ and $S_6$ as functions 
of $\protect \sqrt{q^2}$ for $m_U=300$ GeV and $m_D=300$ GeV.
(i) ${\rm Re}(S_4)$, (ii) ${\rm Im}(S_4)$, (iii) ${\rm Re}(S_6)$, 
(iv) ${\rm Im}(S_6)$.}
\end{figure}
\end{document}